# Extending the La Solubility Limit in $Sr_3Ir_2O_7$ through High-Pressure High-Temperature Synthesis

Cheng Peng, Weiwei Xie*

*Department of Chemistry, Michigan State University, East Lansing, MI 48864 USA*

Corresponding Author: Dr. Weiwei Xie (xieweiwe@msu.edu)

***Abstract***

La-doped bilayer iridates provide an important platform for studying the evolution of the spin-orbit-assisted Mott state under electron doping, but the La solubility achieved by conventional ambient-pressure synthesis is limited. Here, we report the synthesis and physical properties of nominally La-doped $(Sr_{1-x}La_x)_3Ir_2O_7$ ($x$ = 0.05, 0.10, 0.15, and 0.20) prepared using high-pressure high-temperature techniques. Single-crystal X-ray diffraction refinements, supported by scanning electron microscopy-energy-dispersive X-ray spectroscopy (SEM-EDX), reveal significantly enhanced La incorporation, with nominal $x$ = 0.05 and 0.15 corresponding to actual compositions of approximately $(Sr_{0.89}La_{0.11})_3Ir_2O_7$ and $(Sr_{0.77}La_{0.23})_3Ir_2O_7$, respectively. At nominal $x$ = 0.20, the bilayer phase is no longer stabilized and instead transforms into cubic perovskite $Sr_{1-x}La_xIrO_3$. $(Sr_{0.89}La_{0.11})_3Ir_2O_7$ exhibits a ferromagnetic-like transition near 186 K accompanied by magnetic hysteresis and subtle lattice anomalies indicative of spin-lattice coupling. Despite its high electron-doping level, the compound remains strongly insulating, consistent with a heavily doped localized magnetic insulating state distinct from both parent $Sr_3Ir_2O_7$ and ambient-pressure La-doped samples. In contrast, $(Sr_{0.77}La_{0.23})_3Ir_2O_7$ displays metal-like electronic behavior, weakened magnetic order, and enhanced carrier delocalization, although disorder-driven localization persists at low temperatures. These results demonstrate that high-pressure synthesis substantially extends the accessible doping range of bilayer iridates and reveals electronic and magnetic states inaccessible through conventional synthesis routes.

## Introduction

The layered iridates have emerged as an important platform for exploring the interplay of spin-orbit coupling, electron correlations, and lattice degrees of freedom.[1–6] Among them, the bilayer Ruddlesden-Popper iridate $Sr_3Ir_2O_7$ occupies a unique position near the boundary between a spin-orbit-assisted Mott insulator and a correlated metal.[7–10] The delicate balance among Coulomb interactions, crystal-field effects, and strong spin-orbit coupling gives rise to a narrow-gap insulating state derived from the $J_{eff} = 1/2$ electronic configuration.[3,6] Because $Sr_3Ir_2O_7$ lies close to an electronic instability, small perturbations such as chemical substitution, pressure, or strain can strongly modify its magnetic and electronic ground states.[11,12] Electron doping through La substitution on the Sr site has proven to be one of the most effective approaches for tuning the physical properties of $Sr_3Ir_2O_7$.[11,13–16] Previous studies on flux-grown single crystals of $(Sr_{1-x}La_x)_3Ir_2O_7$ have shown that electron doping suppresses the parent antiferromagnetic insulating state and drives the system toward metallic behavior.[17] However, the accessible La concentration is limited by equilibrium crystal-growth conditions, and most studies have been restricted to compositions up to $x \approx 0.08$.[18] Within this narrow doping range, a rich set of phenomena has been observed, including the collapse of long-range antiferromagnetic order, electronic phase separation, persistent magnetic excitations, and unconventional metallic behavior.[11,13–18] These results demonstrate that electron doping does not simply destroy the spin-orbit Mott state but instead produces competing electronic and magnetic phases governed by strong spin-charge-lattice coupling.

High-pressure synthesis provides an alternative strategy for accessing regions of the phase diagram that are inaccessible under ambient-pressure conditions. High pressure has proven particularly effective for stabilizing metastable oxide phases, increasing dopant solubility, and modifying local bonding environments.[19–25] In iridates, pressure can significantly alter Ir-O bond distances, Ir-O-Ir bond angles, and electronic bandwidths, thereby influencing both magnetic exchange interactions and charge transport.[26–28] Unlike conventional flux growth, high-pressure high-temperature synthesis can drive substantially greater incorporation of aliovalent dopants, potentially extending the electron-doping range far beyond previously reported limits.

In this work, we employ high-pressure high-temperature synthesis to investigate La-doped $Sr_3Ir_2O_7$ beyond the doping range accessible through conventional ambient-pressure methods. By combining single-crystal X-ray diffraction, SEM-EDX analysis, magnetization measurements, electrical transport, and variable-temperature structural characterization, we examine how enhanced La incorporation influences the structural, magnetic, and electronic properties of the bilayer iridate. Our study demonstrates that high pressure substantially expands the accessible composition space of La-doped $Sr_3Ir_2O_7$ and provides a unique opportunity to explore emergent phases in a regime that has remained largely inaccessible to previous investigations.

## Experimental Methods

**High-Pressure Synthesis:** Polycrystalline $Sr_2IrO_4$ was first prepared under ambient pressure by thoroughly mixing stoichiometric amounts of $SrCO_3$ and $IrO_2$, followed by pelletization and sequential heat treatments at 900, 1000, and 1100 °C with intermediate grinding steps. The resulting $Sr_2IrO_4$ powder was then mixed with $La_2O_3$ to achieve nominal compositions of $(Sr_{1-x}La_x)_3Ir_2O_7$ ($x$ = 0.05, 0.10, 0.15, and 0.20). The powder mixtures were loaded into Pt capsules and synthesized under high-pressure high-temperature conditions using a Walker-type multi-anvil press at the Lamont-Doherty Earth Observatory, Columbia University. For each synthesis, the Pt capsule was placed inside an $Al_2O_3$ crucible and surrounded by a $LaCrO_3$ heater embedded within a Ceramacast 646 octahedral pressure medium. The assemblies were compressed to 6 GPa, heated to 1400 °C, and held at temperature for 24 h. After synthesis, the samples were cooled to room temperature before releasing pressure. The resulting products are denoted TT-1441 (nominal $x$ = 0.05), BB-1475 (nominal $x$ = 0.10), TT-1442 (nominal $x$ = 0.15), and TT-1445 (nominal $x$ = 0.20).

**Phase Analysis and Chemical Composition:** Phase purity and crystal structure were initially examined by powder X-ray diffraction (PXRD) using a Rigaku MiniFlex 600 diffractometer equipped with Cu $K_\alpha$ radiation ($\lambda$ = 1.5406 Å). Diffraction data were collected at room temperature over the 2θ range of 5-90° with a step size of 0.01° and a scan rate of 1° $min^{-1}$. Lattice parameters and phase fractions were obtained through Le Bail refinements using GSAS-II software suite.[29] Chemical compositions were further evaluated by scanning electron microscopy (SEM) and energy-dispersive X-ray spectroscopy (EDS) using a JEOL JSM-6610LV electron microscope.

**Single-Crystal Structure Determination:** Single-crystal X-ray diffraction measurements were performed on a Rigaku Synergy-S diffractometer equipped with Mo Kα radiation (λ = 0.71073 Å). Data collection and reduction were carried out using the CrysAlisPro software package. Crystal structures were solved by direct methods and refined by full-matrix least-squares techniques on $F^2$ using the SHELXTL program suite.[30,31] Variable-temperature single-crystal diffraction measurements were performed between 100 and 300 K to investigate structural evolution across the magnetic transition.

**Physical Property Measurements:** Temperature- and field-dependent magnetic measurements were performed using a Quantum Design Magnetic Property Measurement System (MPMS3) in the temperature range of 2-300 K under magnetic fields up to 7 T. Electrical transport measurements were carried out using a Quantum Design DynaCool employing a standard four-probe configuration. Platinum wires were attached using silver paste for resistance measurements. Temperature-dependent resistance was measured between 2 and 300 K under applied magnetic fields up to 9 T.

## Results and Discussion

**Phase Identification and Chemical Compositions:** The phase purity and structural evolution of the high-pressure synthesized La-doped $Sr_3Ir_2O_7$ series were examined by powder X-ray diffraction (PXRD), and the results are shown in **Figure 1**. For the nominal $x$ = 0.05 composition, all diffraction peaks can be indexed using the tetragonal $I4/mmm$ structure of the bilayer Ruddlesden-Popper phase. The diffraction pattern is fully consistent with the single-crystal refinement of $Sr_{2.66(3)}La_{0.34}Ir_2O_7$ (discussed below) and no impurity peaks are detected within the resolution of the measurement, indicating the formation of a phase-pure heavily La-doped bilayer iridate. For the nominal $x$ = 0.10 composition, the diffraction pattern remains dominated by the bilayer $I4/mmm$ phase, with peak positions closely matching those of the La-substituted $Sr_3Ir_2O_7$ structure. However, several additional reflections become visible, most notably near $2\theta \approx 28°$ and 44°, indicating the presence of a secondary phase. Although the bilayer phase remains the major component, these impurity peaks suggest that the impurity formation is likely sensitive to synthesis conditions and local chemical environments. Interestingly, for the nominal $x$ = 0.15 composition, the diffraction pattern continues to be dominated by the bilayer phase, while the impurity reflections become noticeably weaker than those observed for $x$ = 0.10. This reduction suggests improved stabilization of the bilayer structure under these synthesis conditions despite the higher La loading. Combined with the single-crystal refinements, which reveal an actual composition of $Sr_{2.30(3)}La_{0.70}Ir_2O_7$, these results demonstrate that high-pressure synthesis enables substantially greater La incorporation than is typically achieved through ambient-pressure crystal growth. A dramatic structural transformation occurs at the nominal $x$ = 0.20 composition. The diffraction pattern of the major phase can no longer be indexed using the bilayer $I4/mmm$ structure and instead is consistent with a cubic perovskite phase. Single-crystal diffraction confirms this phase to be $Sr_{0.4}La_{0.6}IrO_3$ with space group $Pm$-3$m$. The disappearance of the characteristic bilayer reflections and emergence of the cubic perovskite pattern indicate that the $Sr_3Ir_2O_7$ framework is no longer stable at this high La concentration.

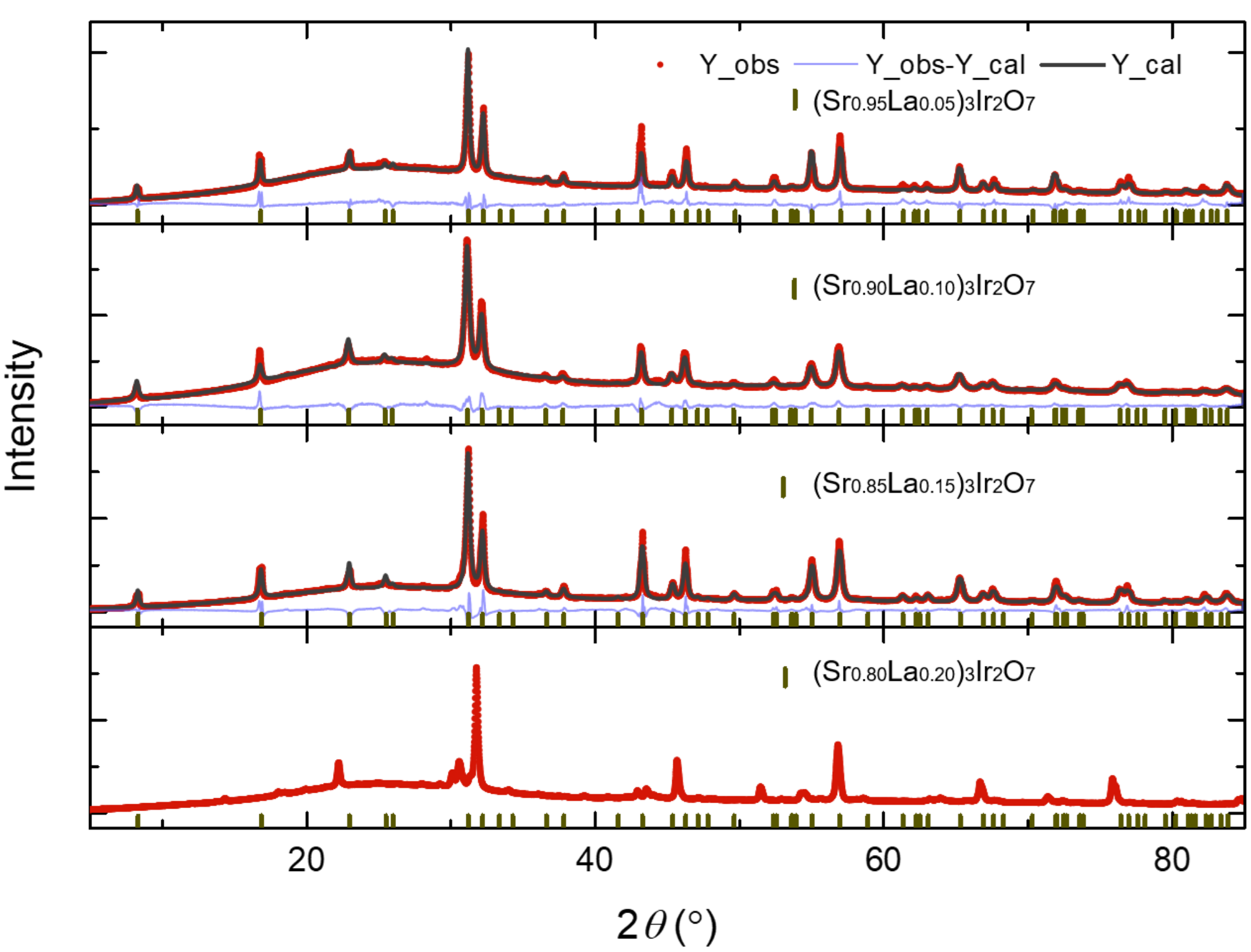

**Figure 1.** Powder X-ray diffraction patterns and Le Bail refinements of high-pressure synthesized La-doped $Sr_3Ir_2O_7$ with nominal compositions $x$ = 0.05, 0.10, 0.15, and 0.20. (Red circles represent experimental data, black lines represent the refinement, blue lines show the difference curve, and green tick marks indicate Bragg reflection positions.)

Single-crystal X-ray diffraction refinements on $x$ = 0.05 and 0.15 compositions at 293 K presented in **Tables 1** and **2** reveal that both compounds adopt the tetragonal *I*4/*mmm* space group characteristic of $Sr_3Ir_2O_7$ shown in **Figure 2*a***. The refinements converge with excellent reliability factors ($R_1$ < 0.03), confirming the high crystallinity of the samples and the suitability of the structural model. A notable result of the structural refinements is the significantly enhanced La incorporation achieved under high-pressure conditions. The nominal $x$ = 0.05 and $x$ = 0.15 samples refine to compositions of $Sr_{2.66(3)}La_{0.34}Ir_2O_7$ and $Sr_{2.30(3)}La_{0.70(3)}Ir_2O_7$, corresponding to actual La concentrations of approximately $x \approx 0.11$ and $x \approx 0.23$, respectively. These values substantially exceed the La concentrations typically reported for ambient-pressure-grown crystals,[11,13–18] demonstrating that high-pressure synthesis effectively expands the accessible electron-doping range in the bilayer iridate system. In addition to the large increase in La content, the average crystal structure remains remarkably robust. The lattice parameters change only slightly from $a$ = 3.9106(6) Å and $c$ = 20.885(4) Å for $Sr_{2.66(3)}La_{0.34}Ir_2O_7$ to $a$ = 3.91562(9) Å and $c$ = 20.8321(9) Å for $Sr_{2.30(3)}La_{0.70(3)}Ir_2O_7$. Consequently, the unit-cell volume remains nearly unchanged at approximately 319.4 Å$^3$ for both compositions. The slight increase of the in-plane lattice parameter accompanied by a small contraction of the $c$-axis suggests anisotropic structural accommodation of the La dopants while preserving the overall bilayer framework. Moreover, La atoms are found to occupy both crystallographically distinct Sr sites, including the rock-salt-layer Sr1 site (2*a*) and the perovskite-layer Sr2 site (4*e*). In both compositions, La preferentially occupies the Sr1 site located within the rock-salt layer, indicating site-selective incorporation of the dopant. This preferential occupation may play an important role in stabilizing the heavily doped bilayer structure under high-pressure conditions. The stability of the Ir-O network, together with the nearly constant unit-cell volume, suggests that the primary effect of La substitution is electronic rather than crystallographic. Therefore, the changes in magnetic and transport properties discussed below are unlikely to originate from a major structural transformation.

**Table 1.** Single-crystal X-ray diffraction refinement data for La-doped $(Sr_{1-x}La_x)_3Ir_2O_7$ ($x$ = 0.05 and 0.15) at 293 K. Values in parentheses represent estimated standard deviations from the refinements.

| **Loading Composition** | $x$ = 0.05 | $x$ = 0.15 |
|---|---|---|
| **Refined composition** | $Sr_{2.66(3)}La_{0.34}Ir_2O_7$ | $Sr_{2.30(3)}La_{0.70(3)}Ir_2O_7$ |
| **Space Group** | *I*4/*mmm* | *I*4/*mmm* |
| ***a* (Å)** | 3.9106(6) | 3.91562(9) |
| ***c* (Å)** | 20.885(4) | 20.8321(9) |

| V (Å$^3$) | 319.4(1) | 319.40(2) |
|---|---|---|
| μ (mm$^{-1}$) | 66.013 | 65.340 |
| F(000) | 659 | 671 |
| θ range (°) | 3.903-36.561 | 3.913-40.614 |
| Data/Restraints/Parameters | 2876/0/24 | 4410/0/23 |
| Indep. Reflections | 281 [$R_{int}$ = 0.0306] | 347 [$R_{int}$ = 0.0289] |
| Reflections (I > 2σ(I)) | 246 | 307 |
| Final R indices | $R_1$(I≥2σ(I)) = 0.0214<br>w$R_2$(I≥2σ(I)) = 0.0456<br>$R_1$(all) = 0.0252<br>w$R_2$(all) = 0.0472 | $R_1$(I≥2σ(I)) = 0.0204<br>w$R_2$(I≥2σ(I)) = 0.0426<br>$R_1$(all) = 0.0262<br>w$R_2$(all) = 0.0438 |
| Goodness-of-fit | 1.131 | 1.121 |
| $\Delta\rho_{max}$ (e-/Å$^3$) | 1.896; -2.802 | 3.504; -2.676 |

**Table 2.** Atomic coordinates and equivalent isotropic displacement parameters ($U_{eq}$) for La-doped $(Sr_{1-x}La_x)_3Ir_2O_7$ ($x$ = 0.05 and 0.15) refined from single-crystal X-ray diffraction collected at 293 K.

| $(Sr_{1-x}La_x)_3Ir_2O_7$ | Atom | Wyck. | *x* | *y* | *z* | Occ. | $U_{eq}$ (Å$^2$) |
|---|---|---|---|---|---|---|---|
| $Sr_{2.66(3)}La_{0.34}Ir_2O_7$ | Sr1 | 2*a* | 0 | 0 | 0 | 0.86(1) | 0.0122(4) |
| | La1 | 2*a* | 0 | 0 | 0 | 0.14(1) | 0.0122(4) |
| | Sr2 | 4*e* | 0 | 0 | 0.18674(6) | 0.93(1) | 0.0097(3) |
| | La2 | 4*e* | 0 | 0 | 0.18674(6) | 0.07(1) | 0.0097(3) |
| | Ir | 4*e* | 0 | 0 | 0.40252(2) | 1 | 0.0051(1) |
| | O1 | 2*b* | 0 | 0 | ½ | 1 | 0.020(2) |
| | O2 | 4*e* | 0 | 0 | 0.3052(3) | 1 | 0.011(1) |
| | O3 | 16*n* | 0 | 0.401(2) | 0.0960(3) | ½ | 0.013(1) |
| $Sr_{2.30(3)}La_{0.70(3)}Ir_2O_7$ | Sr1 | 2*a* | 0 | 0 | 0 | 0.73(1) | 0.0132(3) |
| | La1 | 2*a* | 0 | 0 | 0 | 0.27(1) | 0.0132(3) |
| | Sr2 | 4*e* | 0 | 0 | 0.18649(4) | 0.84(1) | 0.0124(2) |
| | La2 | 4*e* | 0 | 0 | 0.18649(4) | 0.16(1) | 0.0124(2) |
| | Ir | 4*e* | 0 | 0 | 0.40247(2) | 1 | 0.0050(1) |
| | O1 | 2*b* | 0 | 0 | ½ | 1 | 0.030(3) |
| | O2 | 4*e* | 0 | 0 | 0.3046(3) | 1 | 0.013(1) |
| | O3 | 16*n* | 0 | 0.404(2) | 0.0957(3) | ½ | 0.016(1) |

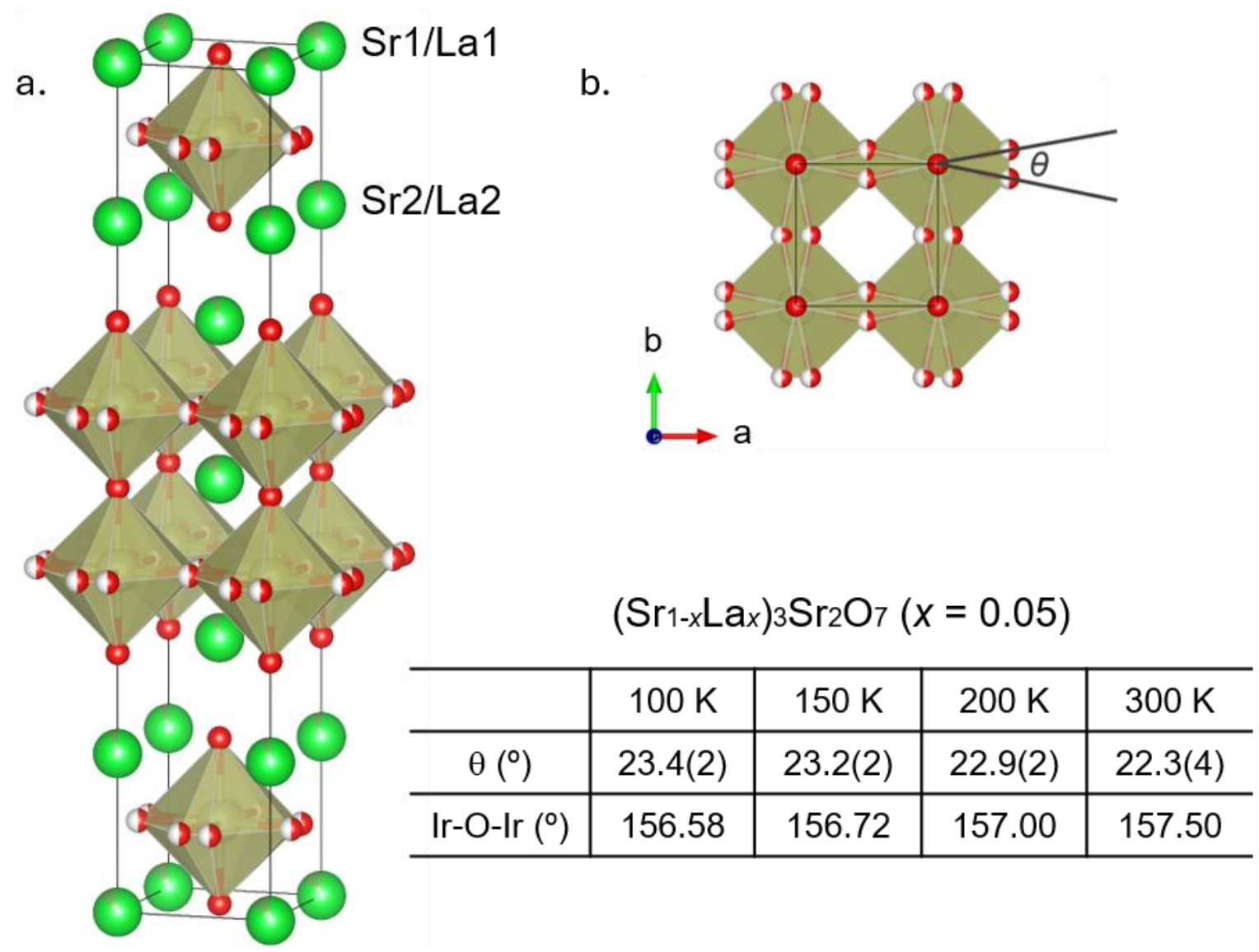

$(Sr_{1-x}La_x)_3Sr_2O_7$ ($x$ = 0.05)

| | 100 K | 150 K | 200 K | 300 K |
|---|---|---|---|---|
| θ (°) | 23.4(2) | 23.2(2) | 22.9(2) | 22.3(4) |
| Ir-O-Ir (°) | 156.58 | 156.72 | 157.00 | 157.50 |

**Figure 2. Crystal structure of the high-pressure synthesized bilayer iridate $(Sr_{1-x}La_x)_3Ir_2O_7$ with tetragonal *I*4/*mmm* symmetry. (*a*)** Crystal structure viewed along the *a*-axis, showing the layered Ruddlesden-Popper framework consisting of corner-sharing $IrO_6$ octahedral bilayers separated by Sr/La-containing rock-salt layers. The two crystallographically distinct alkaline-earth sites, Sr1 and Sr2, are indicated. **(*b*)** Projection of the $IrO_6$ octahedral network along the *c*-axis. The angle $\theta$ denotes the in-plane Ir-O-Ir bond angle, which governs the octahedral rotation and strongly influences the electronic bandwidth and magnetic exchange interactions in the bilayer iridate. The inset table summarizes the temperature dependence of the octahedral rotation angle $\theta$ and the corresponding in-plane Ir-O-Ir bond angle for nominal x=0.05 $(Sr_{1-x}La_x)_3Sr_2O_7$.

**Weak Ferromagnetic-like Transition in $Sr_{2.66(3)}La_{0.34}Ir_2O_7$:** To investigate the magnetic properties of heavily electron-doped $Sr_3Ir_2O_7$, temperature-dependent magnetic susceptibility measurements were performed on $Sr_{2.66(3)}La_{0.34}Ir_2O_7$ (corresponding to approximately 11% La substitution) under applied fields of 100 and 1000 Oe, as shown in **Figure 3*a***. A clear magnetic anomaly is observed between 180 and 190 K, indicating the emergence of a ferromagnetic-like ordered state absent in the parent compound.[7–9] To determine the transition temperature more precisely, the derivative *d*M/*d*T was calculated and is presented in **Figure 3*b***. The minimum in *d*M/*d*T identifies the transition temperature as approximately 186 K. Interestingly, the magnetic transition is most clearly resolved under the lower applied magnetic field of 100 Oe, where the anomaly is observed in both the susceptibility and derivative curves. Upon increasing the applied field to 1000 Oe, the anomaly becomes significantly broader and weaker, indicating that the transition is highly field-sensitive. Such behavior differs from that in a conventional robust ferromagnet, where the transition typically remains well defined under moderate applied fields. Instead, the suppression and broadening of the anomaly suggest the presence of a weak ferromagnetic or canted magnetic state, which is potentially associated with domain effects, spin canting, or field-induced spin reorientation. The observed field dependence demonstrates that the magnetic order in $Sr_{2.66(3)}La_{0.34}Ir_2O_7$ is relatively fragile and likely arises from competing magnetic interactions introduced by heavy electron doping or other couplings.

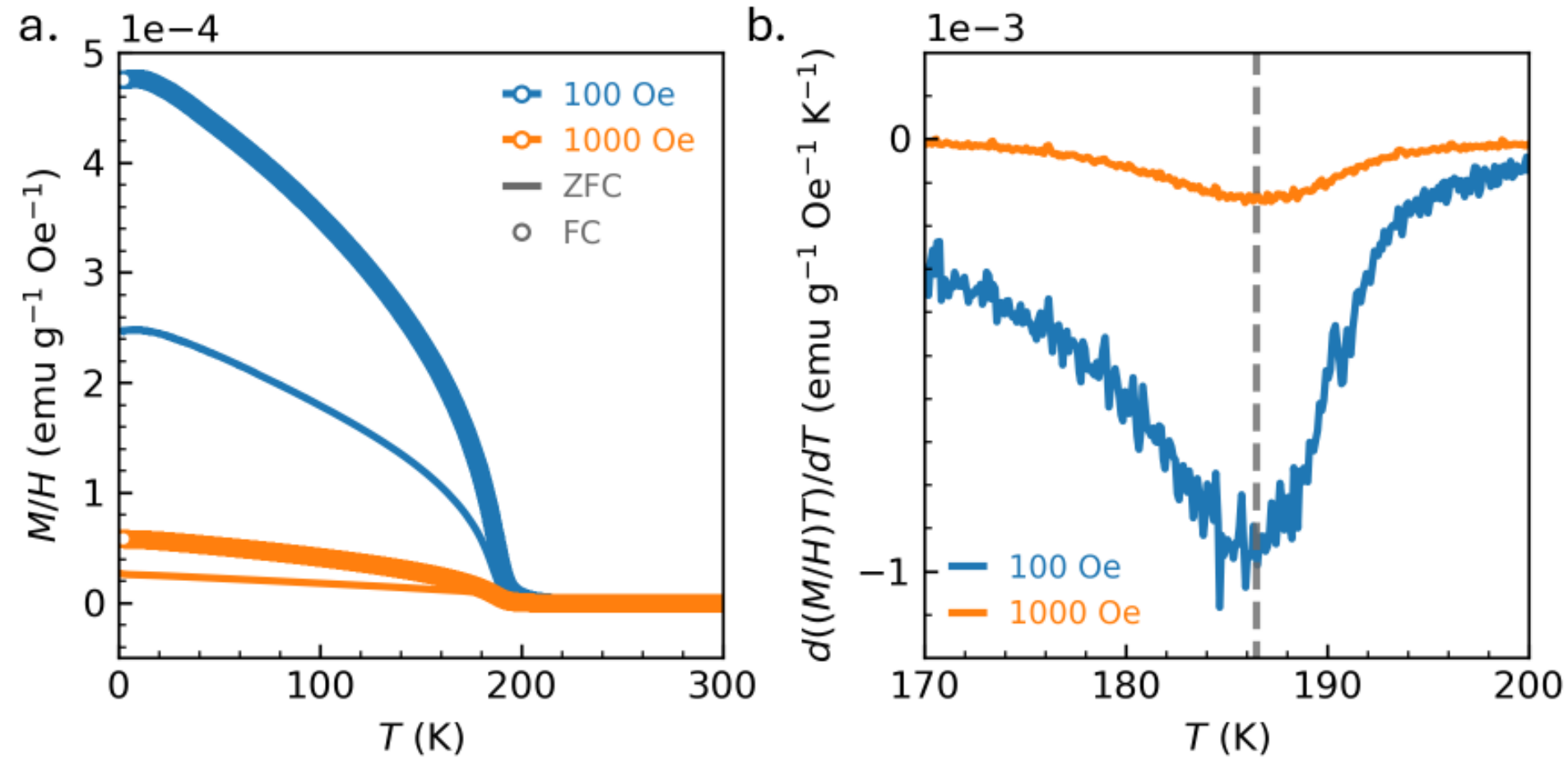

**Figure 3.** **(*a*)** Temperature-dependent magnetic susceptibility of $Sr_{2.66(3)}La_{0.34}Ir_2O_7$ measured under applied fields of 100 and 1000 Oe in zero-field-cooled (ZFC) and field-cooled (FC) modes. **(*b*)** *d*M/*d*T is used to determine the transition temperature.

To further investigate the nature of the magnetic transition, field-dependent magnetization measurements were performed at selected temperatures spanning the ordered and paramagnetic regions, as shown in **Figure 4**. At temperatures well below the transition, the magnetic response is relatively weak. The M(H) curve in **Figure 4*a*** at 2 K is nearly linear, exhibiting little evidence of ferromagnetic order. A small hysteresis loop emerges at 50 K, indicating the presence of a weak ferromagnetic component, although the remanent magnetization and coercive field remain small. As the temperature approaches the magnetic transition, the ferromagnetic-like component becomes significantly enhanced. At 100 K, a pronounced hysteresis loop develops with a remanent magnetization of approximately 0.002 $\mu_B$ per Ir and a coercive field of roughly 1.8 T in **Figure 4*b***. A similarly large hysteretic response is observed at 150 K, indicating that the magnetic state is strongest in the intermediate temperature range below the transition (**Figure 4*c***). At 170 K in **Figure 4*d*** and 180 K in **Figure 4*e***, the hysteresis remains clearly visible but is substantially reduced, consistent with the approach to the transition temperature near 186 K. Above the transition temperature shown in **Figure 4*f***, the ferromagnetic-like response is rapidly suppressed. The hysteresis loop observed at 180 K nearly disappears at 190 K, and the magnetization becomes predominantly linear with field. By 300 K, no measurable hysteresis is detected, confirming the absence of long-range magnetic order in the high-temperature state. The abrupt collapse of the hysteretic response across the transition temperature is consistent with the magnetic anomaly observed in the temperature-dependent susceptibility measurements and supports the existence of a field-sensitive weak ferromagnetic or canted magnetic state in $Sr_{2.66(3)}La_{0.34}Ir_2O_7$.

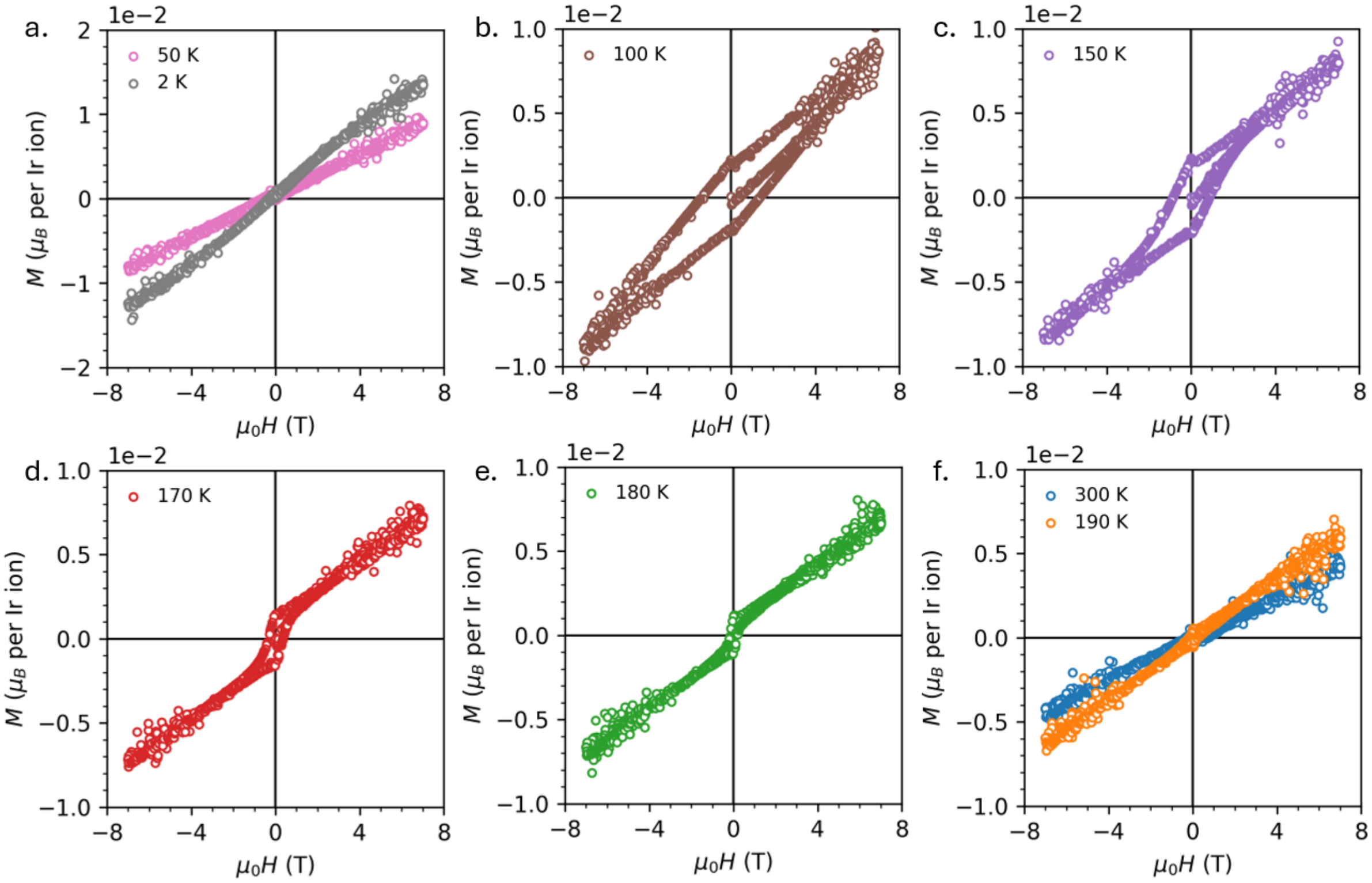

**Figure 4.** Field-dependent magnetization, M(H), of $Sr_{2.66(3)}La_{0.34}Ir_2O_7$ measured at selected temperatures. ***(a)*** M(H) curves at 2 and 50 K. ***(b-e)*** M(H) curves at 100, 150, 170 and 180 K showing clear hysteresis below the magnetic transition. ***(f)*** M(H) curves at 190 and 300 K, where the ferromagnetic-like response is strongly suppressed. The largest hysteresis is observed between 100 and 150 K, while the response becomes nearly linear above the transition temperature of approximately 186 K.

**Localized Charge Transport in $Sr_{2.66(3)}La_{0.34}Ir_2O_7$:** The electrical resistance of $Sr_{2.66(3)}La_{0.34}Ir_2O_7$ was measured under applied magnetic fields of 0 and 1 T, as shown in **Figure 5*a***. The resistance increases rapidly upon cooling, indicating strongly insulating behavior over the entire measured temperature range. The 0 T and 1 T curves nearly overlap, demonstrating that the transport is essentially field-independent. This absence of measurable magnetoresistance suggests that the charge carriers remain strongly localized and that the ferromagnetic-like magnetic transition does not significantly modify the dominant conduction pathway. To examine possible transport anomalies near the magnetic transition, *d*R/*d*T was calculated around 170-200 K, as shown in **Figure 5*b***. Only a subtle change in slope is observed near 186 K, consistent with weak coupling between the magnetic transition and charge transport. Arrhenius analysis of the resistance reveals two thermally activated regimes separated by the magnetic transition (**Figure 5*c***). Above 186 K, the activation energy is approximately 68.40(1) meV, whereas below the transition it increases to 82.42(2) meV. The larger activation energy below the magnetic transition indicates that the ordered magnetic state further enhances carrier localization. These results show that $Sr_{2.66(3)}La_{0.34}Ir_2O_7$ is a heavily electron-doped magnetic insulator in which spin ordering is coupled to carrier localization but does not produce field-tunable transport.

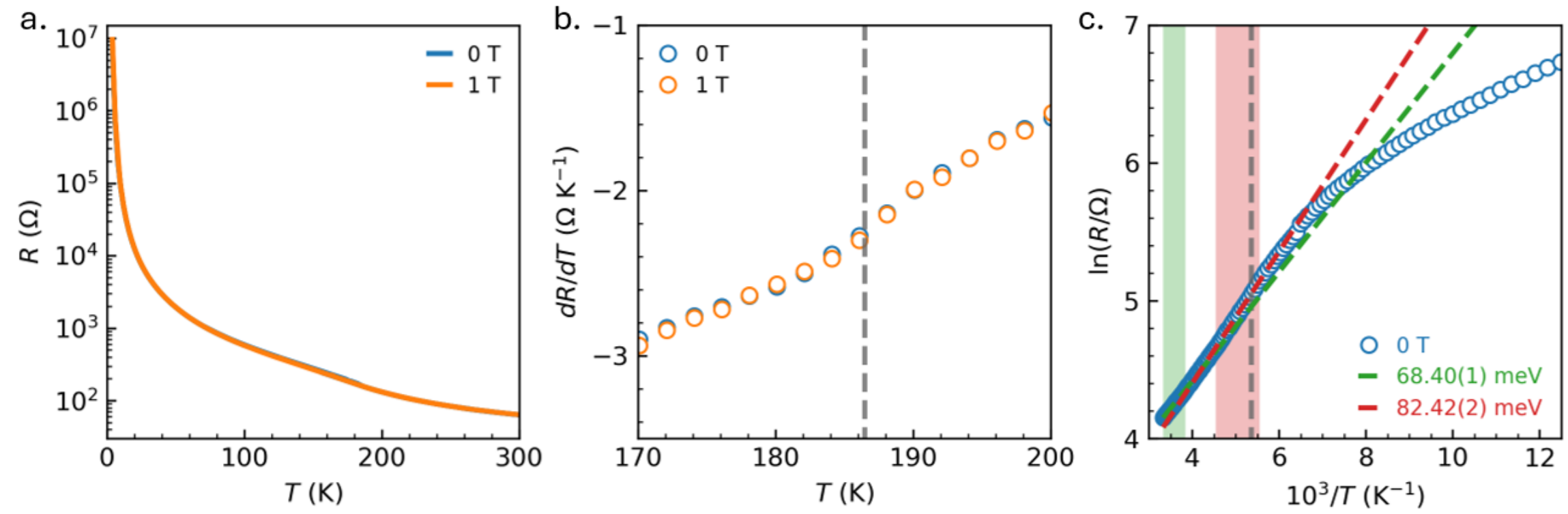

**Figure 5.** Electrical transport of $Sr_{2.66(3)}La_{0.34}Ir_2O_7$. **(*a*)** Temperature-dependent resistance measured under 0 and 1 T, showing strongly insulating behavior with negligible field dependence. **(*b*)** *d*R/*d*T near the magnetic transition region, showing only a weak slope change around 186 K. **(*c*)** Arrhenius analysis, ln(*R*) versus $10^3$/T, showing two thermally activated regimes with activation energies of 68.40(1) meV above the transition and 82.42(2) meV below the transition.

**Weak Spin-Lattice Coupling Observed in $Sr_{2.66(3)}La_{0.34}Ir_2O_7$ Across the Magnetic Transition:** To investigate the relationship between the crystal structure and the ferromagnetic-like transition observed near 186 K, temperature-dependent single-crystal X-ray diffraction measurements were performed on $Sr_{2.66(3)}La_{0.34}Ir_2O_7$ between 100 and 300 K. At all measured temperatures, the structure remains tetragonal with space group *I*4/*mmm*, indicating that the magnetic transition is not accompanied by a crystallographic symmetry-breaking structural transition. The refined lattice parameters and unit-cell volumes are summarized in **Tables S3 and S4**. As expected, the unit-cell volume increases monotonically with increasing temperature due to thermal expansion. However, the thermal expansion is not uniform throughout the measured temperature range. Between 100 and 150 K, the volume increases by only 0.187 $Å^3$, whereas substantially larger increases of 0.472 and 0.458 $Å^3$ are observed between 150-200 K and 200-300 K, respectively. The enhanced expansion occurring in the 150-200 K interval coincides with the magnetic transition temperature of approximately 186 K, suggesting a subtle magnetoelastic anomaly associated with the onset of magnetic order. Although no discontinuity or symmetry change is observed, the anomalous thermal expansion indicates that the magnetic state is coupled to the lattice. To further probe the origin of this anomaly, the local geometry of the $IrO_6$ octahedral framework was examined, as shown in **Figure 2*b***. The in-plane Ir-O-Ir bond angle increases systematically from 156.58° at 100 K to 156.72° at 150 K, 157.00° at 200 K, and 157.50° at 300 K, corresponding to a gradual reduction of the octahedral rotation with increasing temperature. Notably, the largest change in the bond angle occurs across the magnetic transition region. Because the Ir-O-Ir bond angle directly controls both the electronic bandwidth and magnetic exchange interactions in layered iridates, the observed evolution indicates that the magnetic transition is accompanied by a subtle modification of the $IrO_6$ network.[26–28] Below the transition, the smaller Ir-O-Ir angle corresponds to enhanced octahedral rotation and reduced orbital overlap, consistent with the larger activation energy extracted from transport measurements. The combined evolution of the unit-cell volume and Ir-O-Ir bond angle provides direct evidence for weak spin-lattice coupling in $Sr_{2.66(3)}La_{0.34}Ir_2O_7$. While the structural response is modest and does not involve symmetry breaking, the coincidence of the lattice anomalies with the magnetic transition demonstrates that the ferromagnetic-like state is coupled to distortions of the $IrO_6$ framework. Such coupling may contribute to the enhanced carrier localization observed below the transition and highlights the important role of lattice degrees of freedom in stabilizing the heavily electron-doped magnetic insulating state.

**Suppression of the Magnetic Insulating State in $Sr_{2.30(3)}La_{0.70}Ir_2O_7$:** To understand the evolution of the magnetic insulating state with increasing electron doping, the magnetic and transport properties of $Sr_{2.30(3)}La_{0.70}Ir_2O_7$ were investigated and compared with those of $Sr_{2.66(3)}La_{0.34}Ir_2O_7$. As shown in **Figure 6*a***, the temperature-dependent magnetic susceptibility exhibits predominantly paramagnetic behavior over the entire measured temperature range. In contrast to $Sr_{2.66(3)}La_{0.34}Ir_2O_7$, no ferromagnetic-like transition is observed near 186 K, indicating that the magnetic state stabilized in the lower-doped sample is completely suppressed at higher La concentrations. A weak low-temperature upturn is present below approximately 30 K and is likely associated with the minor impurity phase detected in the powder X-ray diffraction measurements. The field-dependent magnetization measurements further support the absence of long-range magnetic order. As shown in **Figure 6*b***, the M(H) curves are nearly linear at 300, 150, and 50 K, consistent with paramagnetic behavior. At 20 K, a slight deviation from linearity and a weak hysteretic component are observed, which likely originate from the impurity phase rather than the intrinsic $Sr_{2.30(3)}La_{0.70}Ir_2O_7$ phase. Unlike $Sr_{2.66(3)}La_{0.34}Ir_2O_7$, no substantial remanent magnetization or coercive field is detected. The temperature-dependent resistance measurements reveal a markedly different electronic state compared with the heavily doped magnetic insulator. As shown in **Figure 6*c***, the resistance decreases with decreasing temperature at high temperatures, indicative of metallic or semimetallic transport. Upon further cooling, a broad minimum develops and the resistance increases at low temperatures, suggesting a crossover to a weakly localized state. In contrast to $Sr_{2.66(3)}La_{0.34}Ir_2O_7$, which remains strongly insulating throughout the measured temperature range, $Sr_{2.30(3)}La_{0.70}Ir_2O_7$ exhibits substantially enhanced carrier delocalization. These results indicate that increasing La incorporation suppresses the ferromagnetic-like insulating state and drives the system toward a more itinerant electronic regime, although complete metallic behavior is not achieved due to persistent disorder and localization effects.

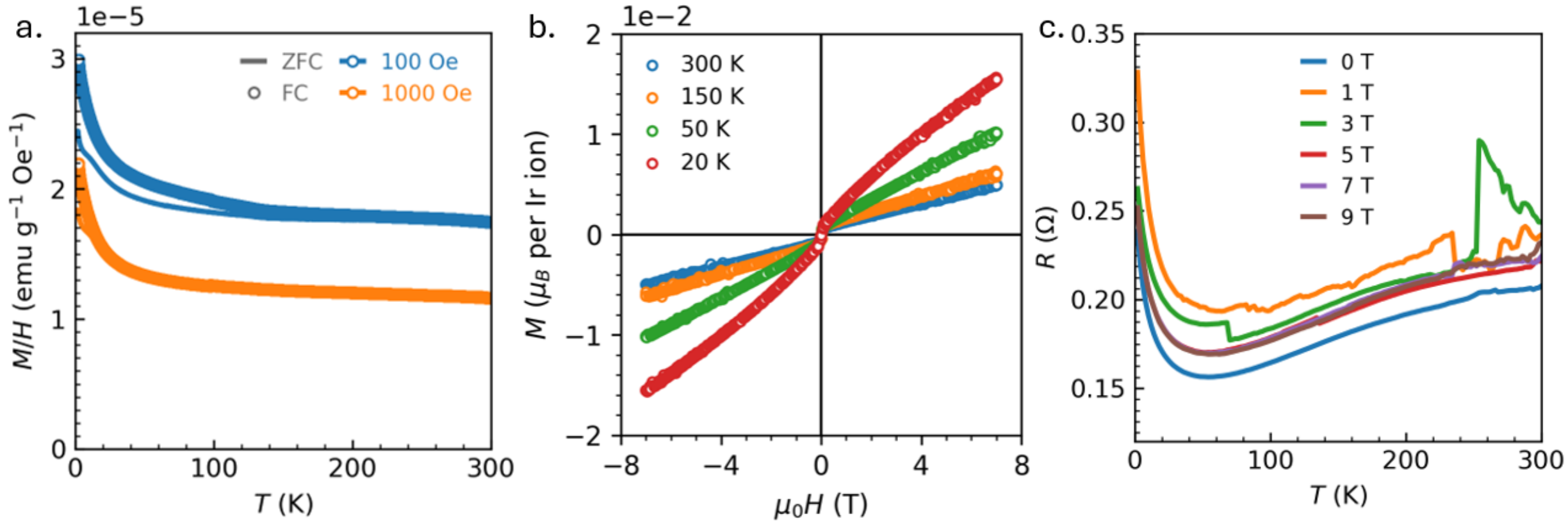

**Figure 6.** Physical properties of $Sr_{2.30(3)}La_{0.70}Ir_2O_7$. **(*a*)** Temperature-dependent magnetic susceptibility measured under applied fields of 100 and 1000 Oe in zero-field-cooled (ZFC) and field-cooled (FC) modes. **(*b*)** Field-dependent magnetization measured at selected temperatures, showing predominantly paramagnetic behavior. **(*c*)** Temperature-dependent resistance measured at magnetic fields from 0 to 9 T.

## Conclusion

High-pressure high-temperature synthesis significantly extends the accessible La-doping range in the bilayer iridate $Sr_3Ir_2O_7$ beyond that achieved by conventional ambient-pressure methods. Single-crystal X-ray diffraction reveals that nominal compositions of $x$ = 0.05 and 0.15 correspond to actual compositions of $Sr_{2.66(3)}La_{0.34}Ir_2O_7$ and $Sr_{2.30(3)}La_{0.70}Ir_2O_7$, respectively, while further La loading drives a structural transformation to cubic perovskite $Sr_{0.4}La_{0.6}IrO_3$. $Sr_{2.66(3)}La_{0.34}Ir_2O_7$ exhibits an unexpected ferromagnetic-like transition near 186 K together with strongly insulating transport and weak spin-lattice coupling, establishing a heavily electron-doped magnetic insulating state distinct from both parent $Sr_3Ir_2O_7$ and previously reported ambient-pressure La-doped samples. In contrast, $Sr_{2.30(3)}La_{0.70}Ir_2O_7$ shows paramagnetic behavior and enhanced carrier delocalization, indicating suppression of the magnetic insulating state with increasing electron doping. These results demonstrate that high-pressure synthesis provides access to previously unexplored regions of the electron-doped phase diagram and reveals emergent magnetic and electronic states inaccessible through conventional synthesis routes.

## Supporting Information

Supporting Information is available free of charge on the ACS website.

The crystal structure and refinement of La-doped $(Sr_{1-x}La_x)_3Ir_2O_7$ ($x$ = 0.20); Atomic coordinates of La-doped $(Sr_{1-x}La_x)_3Ir_2O_7$ ($x$ = 0.20); The crystal structure and refinement of La-doped $(Sr_{1-x}La_x)_3Ir_2O_7$ ($x$ = 0.05) from 100 K to 300 K; Atomic coordinates of La-doped $(Sr_{1-x}La_x)_3Ir_2O_7$ ($x$ = 0.05) from 100 K to 300 K;SEM-EDX results of La-doped $(Sr_{1-x}La_x)_3Ir_2O_7$ ($x$ = 0.05, 0.10, 0.15, and 0.20)

## Conflicts of Interest

The authors declare no competing interest.

## Acknowledgments

C.P. and W.X. appreciate Haozhe Wang's previous work on this project. The work at MSU was supported by NSF-DMR-2422361. C.P. was also supported by the S.M.A.R.T. fellowship from Michigan State University.

## Table of Contents

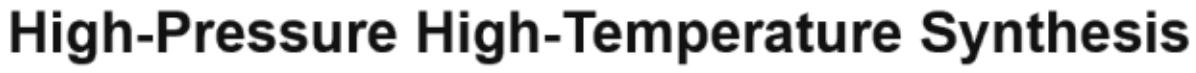

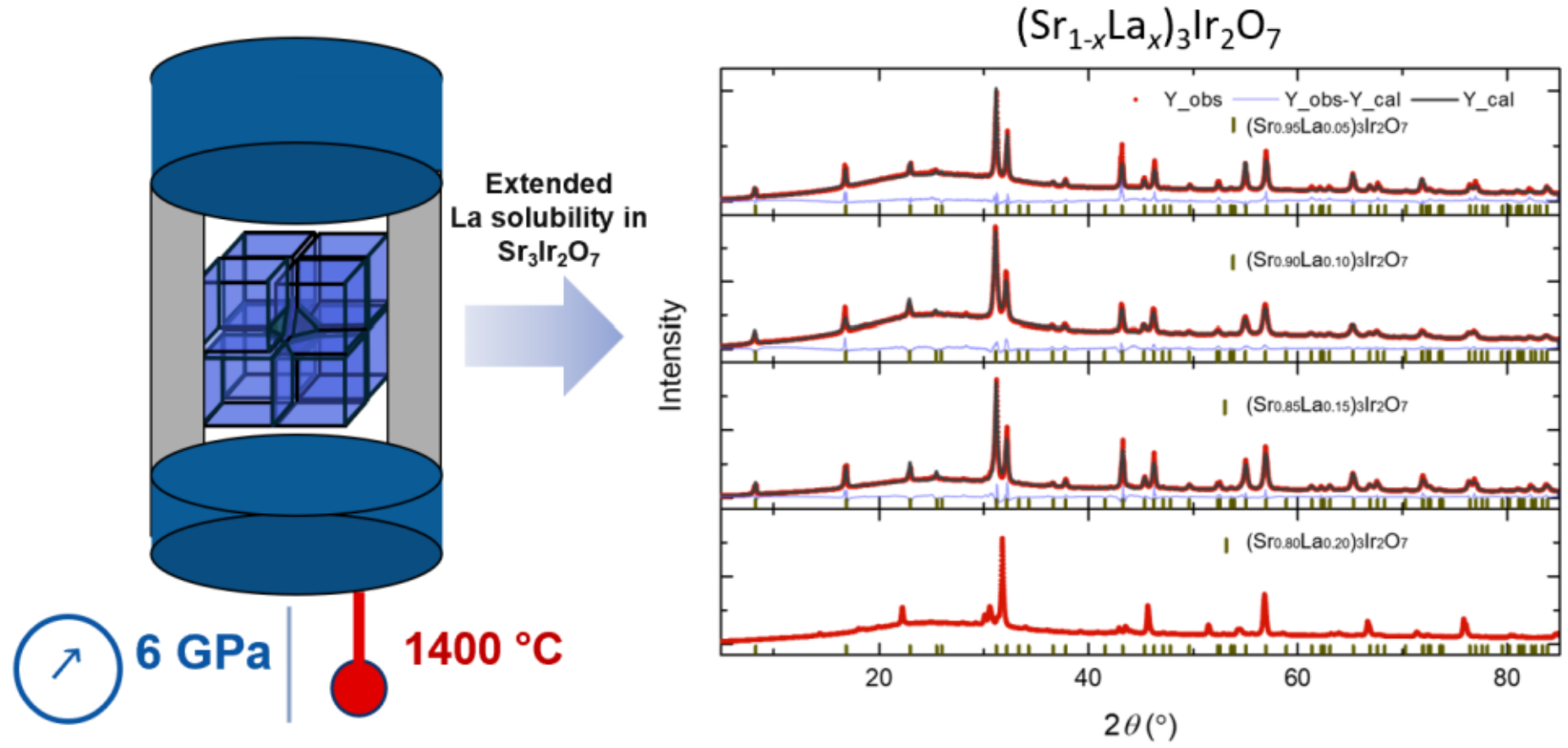

High-pressure, high-temperature synthesis extends La incorporation in bilayer Sr3Ir2O7, stabilizing a localized ferromagnetic-like insulating state before higher La loading drives a more itinerant regime and, eventually, a cubic perovskite phase.

## Supplementary Information

# Extending the La Solubility Limit in $Sr_3Ir_2O_7$ through High-Pressure High-Temperature Synthesis

Cheng Peng, Weiwei Xie*

*Department of Chemistry, Michigan State University, East Lansing, MI 48864 USA*

Corresponding Author: Dr. Weiwei Xie (xieweiwe@msu.edu)

### Table of Contents

**Table S1.** Single crystal X-ray diffraction refinement for nominal $(Sr_{0.8}La_{0.2})_3Ir_2O_7$ at 293 K.

| **Loading Composition** | $x = 0.20$ |
|---|---|
| **Refined composition** | $Sr_{0.40(7)}La_{0.60(7)}IrO_3$ |
| **Space Group** | *Pm*-3*m* |
| ***a*** **(Å)** | 3.9515(5) |
| **V (Å$^3$)** | 61.70(2) |
| **μ (mm$^{-1}$)** | 92.052 |
| **F(000)** | 196 |
| **θ range (º)** | 5.160-48.333 |
| **Data/Restraints/Parameters** | 2786/0/7 |
| **Indep. Reflections** | 89 [$R_{int}$ = 0.0509] |
| **Reflections (I > 2σ(I))** | 89 |
| **Final R indices** | $R_1$(I≥2σ(I)) = 0.0379<br>w$R_2$(I≥2σ(I)) = 0.0875<br>$R_1$(all) = 0.0379<br>w$R_2$(all) = 0.0875 |
| **Goodness-of-fit** | 1.179 |
| **$\Delta\rho_{max}$ (e-/Å$^3$)** | 4.041; -4.992 |

**Table S2.** Atomic coordinates and equivalent isotropic displacement parameters for nominal $(Sr_{0.8}La_{0.2})_3Ir_2O_7$.

| | **Atom** | **Wyck.** | ***x*** | ***y*** | ***z*** | **Occ.** | **$U_{eq}$ (Å$^2$)** |
|---|---|---|---|---|---|---|---|
| $Sr_{0.40(7)}La_{0.60}IrO_3$ | Sr1 | 1*b* | ½ | ½ | ½ | 0.40(7) | 0.025(1) |
| | La1 | 1*b* | ½ | ½ | ½ | 0.60(7) | 0.025(1) |
| | Ir | 1*a* | 0 | 0 | 0 | 1 | 0.0031(4) |
| | O | 3*d* | 0 | 0 | ½ | 1 | 0.13(2) |

**Table S3.** Single crystal X-ray diffraction refinement for the nominal x=0.05 sample, $Sr_{2.66(3)}La_{0.34}Ir_2O_7$ at 100 K, 150 K, 200 K and 300 K.

| **Temperature** | 100 K | 150 K | 200 K | 300 K |
|---|---|---|---|---|
| **Space Group** | *I*4/*mmm* | *I*4/*mmm* | *I*4/*mmm* | *I*4/*mmm* |
| ***a*** **(Å)** | 3.90117(5) | 3.90232(5) | 3.90452(5) | 3.90864(6) |
| ***c*** **(Å)** | 20.8666(6) | 20.8665(6) | 20.8740(6) | 20.8600(7) |
| **V (Å$^3$)** | 317.57(1) | 317.76(1) | 318.23(1) | 318.69(1) |
| **μ (mm$^{-1}$)** | 66.245 | 66.206 | 66.108 | 66.013 |
| **F(000)** | 659 | 659 | 659 | 659 |
| **θ range (°)** | 3.906-41.129 | 3.906-41.117 | 3.905-40.952 | 3.907-41.034 |
| **Data/Restraints/Parameters** | 4684/0/23 | 4604/0/23 | 4824/0/23 | 4577/0/23 |
| **Indep. Reflections** | 368 [$R_{int}$ = 0.0170] | 367 [$R_{int}$ = 0.0161] | 367 [$R_{int}$ = 0.0149] | 368 [$R_{int}$ = 0.0186] |
| **Reflections (I > 2σ(I))** | 336 | 337 | 329 | 339 |
| **Final R indices** | $R_1$(I≥2σ(I)) = 0.0115<br>w$R_2$(I≥2σ(I)) = 0.0254<br>$R_1$(all) = 0.0142<br>w$R_2$(all) = 0.0259 | $R_1$(I≥2σ(I)) = 0.0123<br>w$R_2$(I≥2σ(I)) = 0.0269<br>$R_1$(all) = 0.0146<br>w$R_2$(all) = 0.0274 | $R_1$(I≥2σ(I)) = 0.0118<br>w$R_2$(I≥2σ(I)) = 0.0280<br>$R_1$(all) = 0.0138<br>w$R_2$(all) = 0.0284 | $R_1$(I≥2σ(I)) = 0.0155<br>w$R_2$(I≥2σ(I)) = 0.0387<br>$R_1$(all) = 0.0183<br>w$R_2$(all) = 0.0394 |
| **Goodness-of-fit** | 1.130 | 1.123 | 1.077 | 1.145 |
| **$\Delta\rho_{max}$ (e-/Å$^3$)** | 2.062; -1.931 | 2.186; -1.388 | 2.0166; -1.315 | 2.318; -1.674 |

**Table S4.** Atomic coordinates and equivalent isotropic displacement parameters ($U_{eq}$) for the nominal x=0.05 sample, $Sr_{2.66(3)}La_{0.34}Ir_2O_7$ at 100 K, 150 K, 200 K and 300 K.

| $(Sr_{0.95}La_{0.05})_3Ir_2O_7$ | **Atom** | **Wyck.** | ***x*** | ***y*** | ***z*** | **Occ.** | $U_{eq}$ **(Å²)** |
|---|---|---|---|---|---|---|---|
| 100 K | Sr1 | 2*a* | 0 | 0 | 0 | 0.86(1) | 0.0064(1) |
| | La1 | 2*a* | 0 | 0 | 0 | 0.14(1) | 0.0064(1) |
| | Sr2 | 4*e* | 0 | 0 | 0.18683(2) | 0.92(1) | 0.0056(1) |
| | La2 | 4*e* | 0 | 0 | 0.18683(2) | 0.08(1) | 0.0056(1) |
| | Ir | 4*e* | 0 | 0 | 0.40251(2) | 1 | 0.0022(1) |
| | O1 | 2*b* | 0 | 0 | ½ | 1 | 0.0149(9) |
| | O2 | 4*e* | 0 | 0 | 0.3051(1) | 1 | 0.0064(5) |
| | O3 | 16*n* | 0 | 0.3967(8) | 0.0959(1) | ½ | 0.0072(5) |
| 150 K | Sr1 | 2*a* | 0 | 0 | 0 | 0.87(1) | 0.0075(1) |
| | La1 | 2*a* | 0 | 0 | 0 | 0.13(1) | 0.0075(1) |
| | Sr2 | 4*e* | 0 | 0 | 0.18682(2) | 0.92(1) | 0.0063(1) |
| | La2 | 4*e* | 0 | 0 | 0.18682(2) | 0.08(1) | 0.0063(1) |
| | Ir | 4*e* | 0 | 0 | 0.40252(2) | 1 | 0.0026(1) |
| | O1 | 2*b* | 0 | 0 | ½ | 1 | 0.0154(9) |
| | O2 | 4*e* | 0 | 0 | 0.3050(1) | 1 | 0.0063(5) |
| | O3 | 16*n* | 0 | 0.3974(8) | 0.0958(1) | ½ | 0.0088(5) |
| 200 K | Sr1 | 2*a* | 0 | 0 | 0 | 0.86(1) | 0.0087(2) |
| | La1 | 2*a* | 0 | 0 | 0 | 0.14(1) | 0.0087(2) |
| | Sr2 | 4*e* | 0 | 0 | 0.18680(3) | 0.92(1) | 0.0073(1) |
| | La2 | 4*e* | 0 | 0 | 0.18680(3) | 0.08(1) | 0.0073(1) |
| | Ir | 4*e* | 0 | 0 | 0.40252(2) | 1 | 0.0030(1) |
| | O1 | 2*b* | 0 | 0 | ½ | 1 | 0.0183(11) |
| | O2 | 4*e* | 0 | 0 | 0.3052(1) | 1 | 0.0080(6) |
| | O3 | 16*n* | 0 | 0.3988(9) | 0.0959(2) | ½ | 0.0098(5) |
| 300 K | Sr1 | 2*a* | 0 | 0 | 0 | 0.86(1) | 0.0109(2) |
| | La1 | 2*a* | 0 | 0 | 0 | 0.14(1) | 0.0109(2) |
| | Sr2 | 4*e* | 0 | 0 | 0.18671(3) | 0.92(1) | 0.0096(2) |
| | La2 | 4*e* | 0 | 0 | 0.18671(3) | 0.08(1) | 0.0096(2) |
| | Ir | 4*e* | 0 | 0 | 0.40249(2) | 1 | 0.0040(1) |
| | O1 | 2*b* | 0 | 0 | ½ | 1 | 0.019(2) |
| | O2 | 4*e* | 0 | 0 | 0.3054(2) | 1 | 0.011(1) |
| | O3 | 16*n* | 0 | 0.401(1) | 0.0961(2) | ½ | 0.012(1) |

**Figure S1.** SEM-EDX results for La-doped $(Sr_{1-x}La_x)_3Ir_2O_7$

***x* = 0.05**

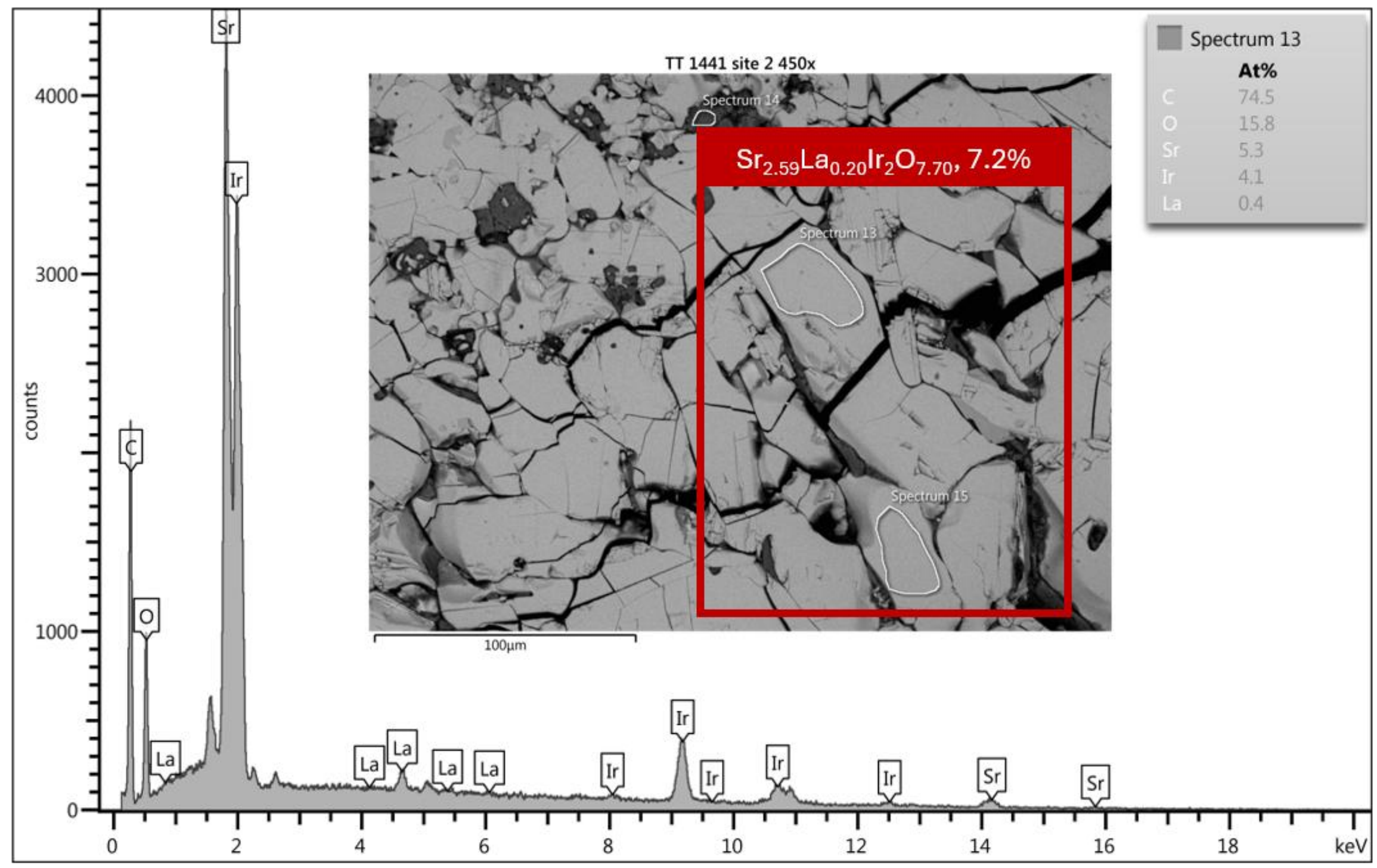

***x* = 0.10**

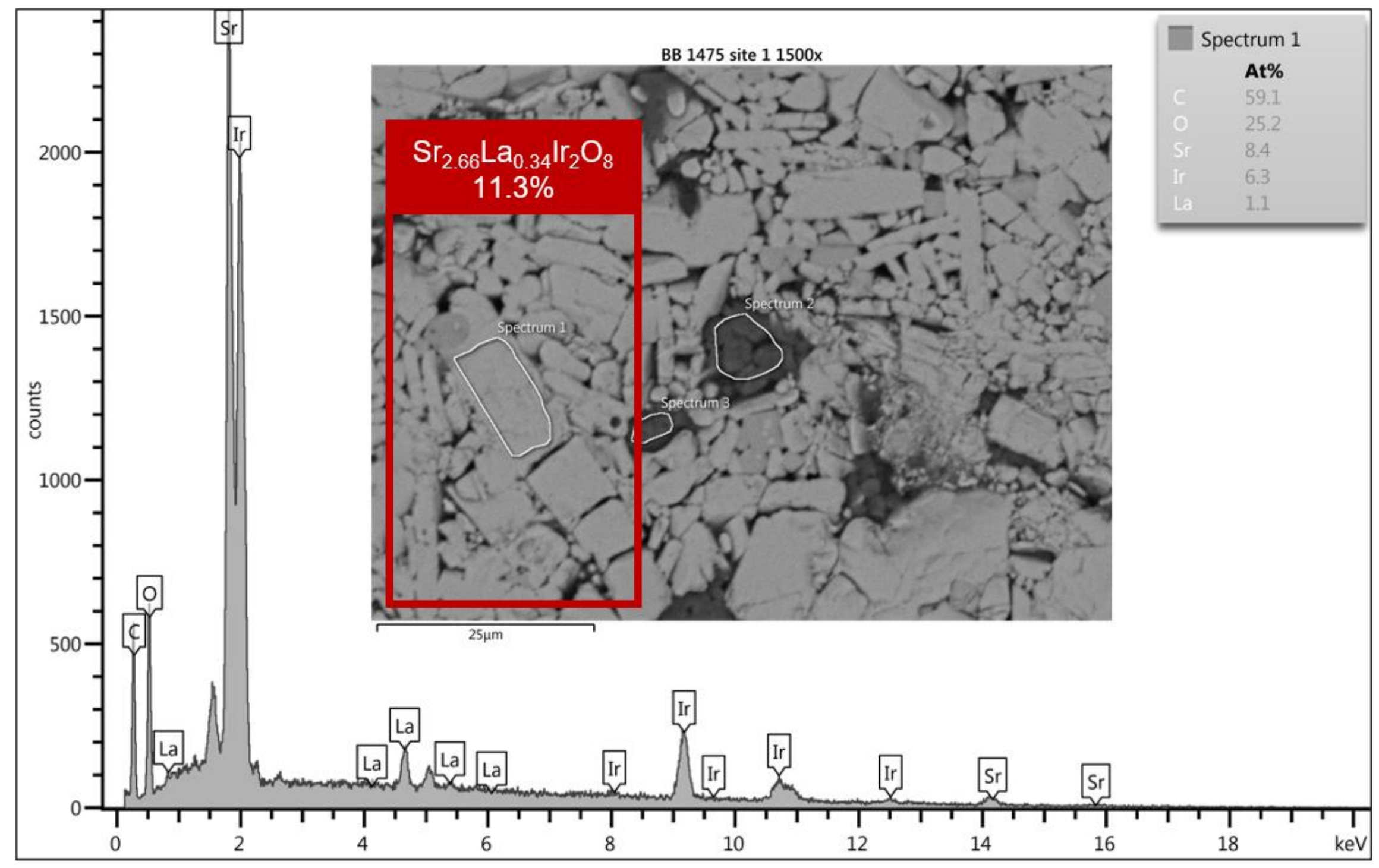

***x* = 0.15**

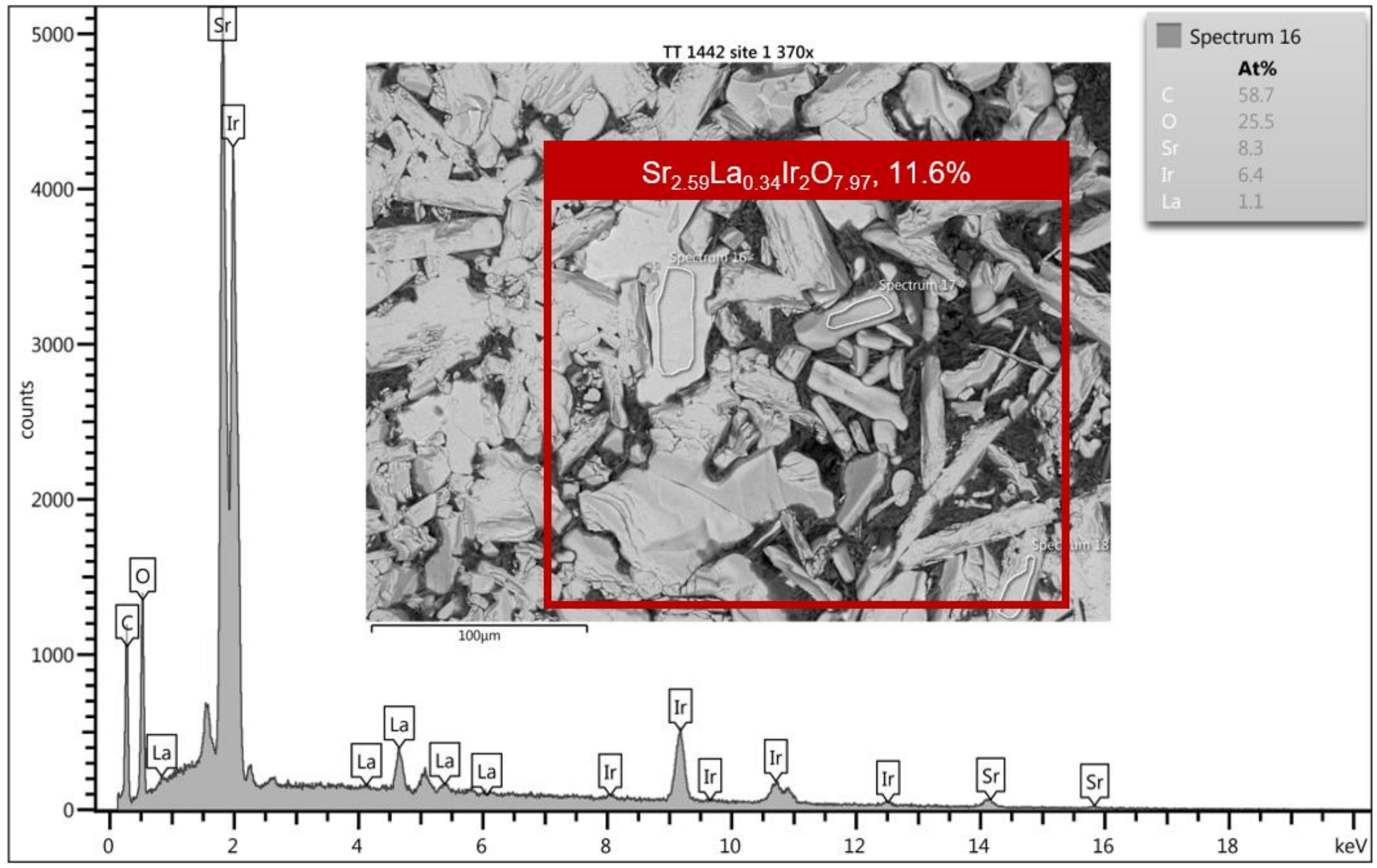

TT 1442 site 1 370x
$Sr_{2.59}La_{0.34}Ir_2O_{7.97}$, 11.6%
Spectrum 16
Spectrum 17
100µm
Spectrum 16
At%
C 58.7
O 25.5
Sr 8.3
Ir 6.4
La 1.1
counts
keV

***x* = 0.20**

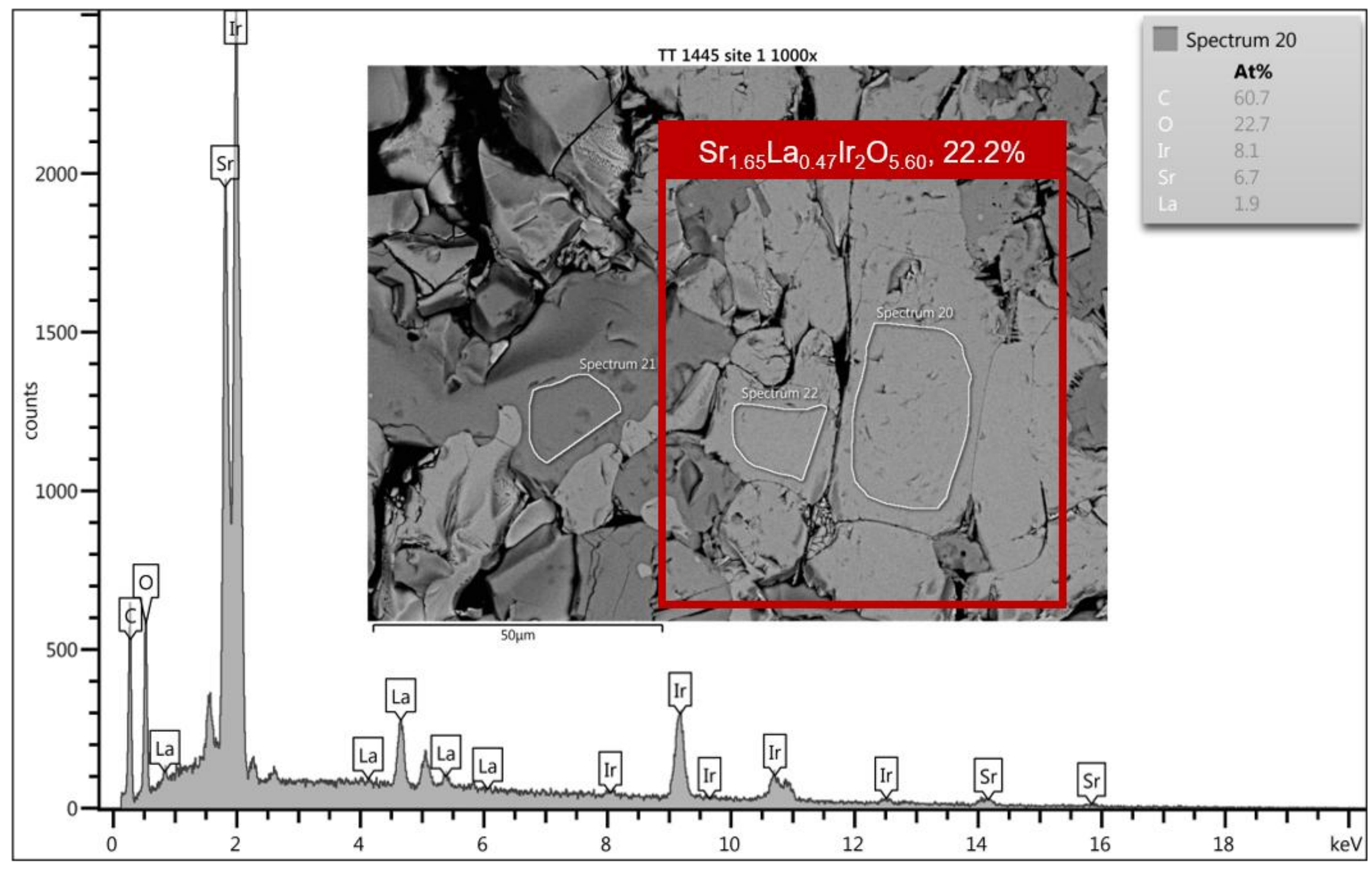

TT 1445 site 1 1000x
$Sr_{1.65}La_{0.47}Ir_2O_{5.60}$, 22.2%
Spectrum 20
Spectrum 21
Spectrum 22
50µm
Spectrum 20
At%
C 60.7
O 22.7
Ir 8.1
Sr 6.7
La 1.9
counts
keV

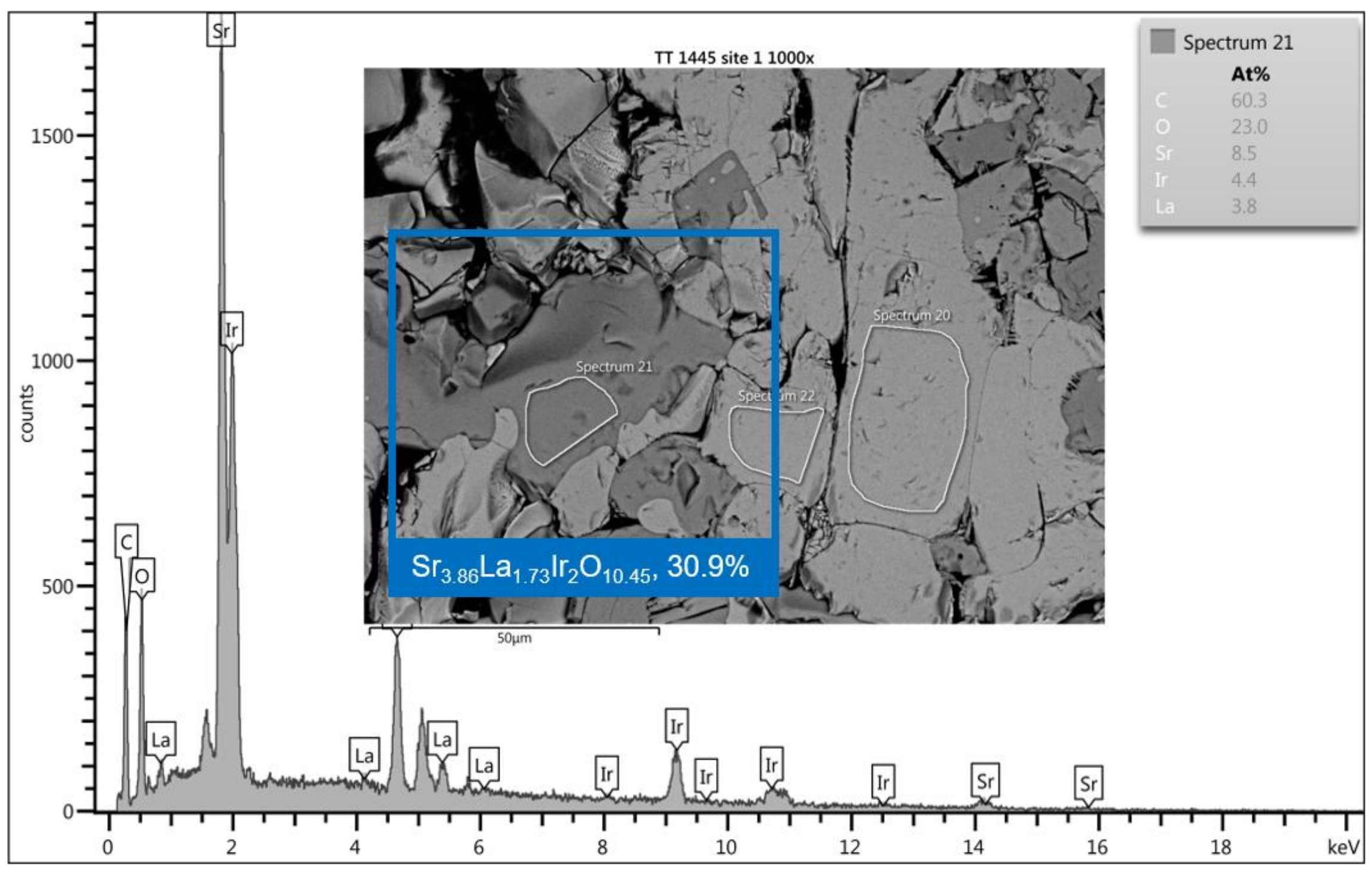

TT 1445 site 1 1000x
Spectrum 21
At%
C 60.3
O 23.0
Sr 8.5
Ir 4.4
La 3.8
Spectrum 20
Spectrum 21
Spectrum 22
$Sr_{3.86}La_{1.73}Ir_2O_{10.45}$, 30.9%
50µm
counts
keV

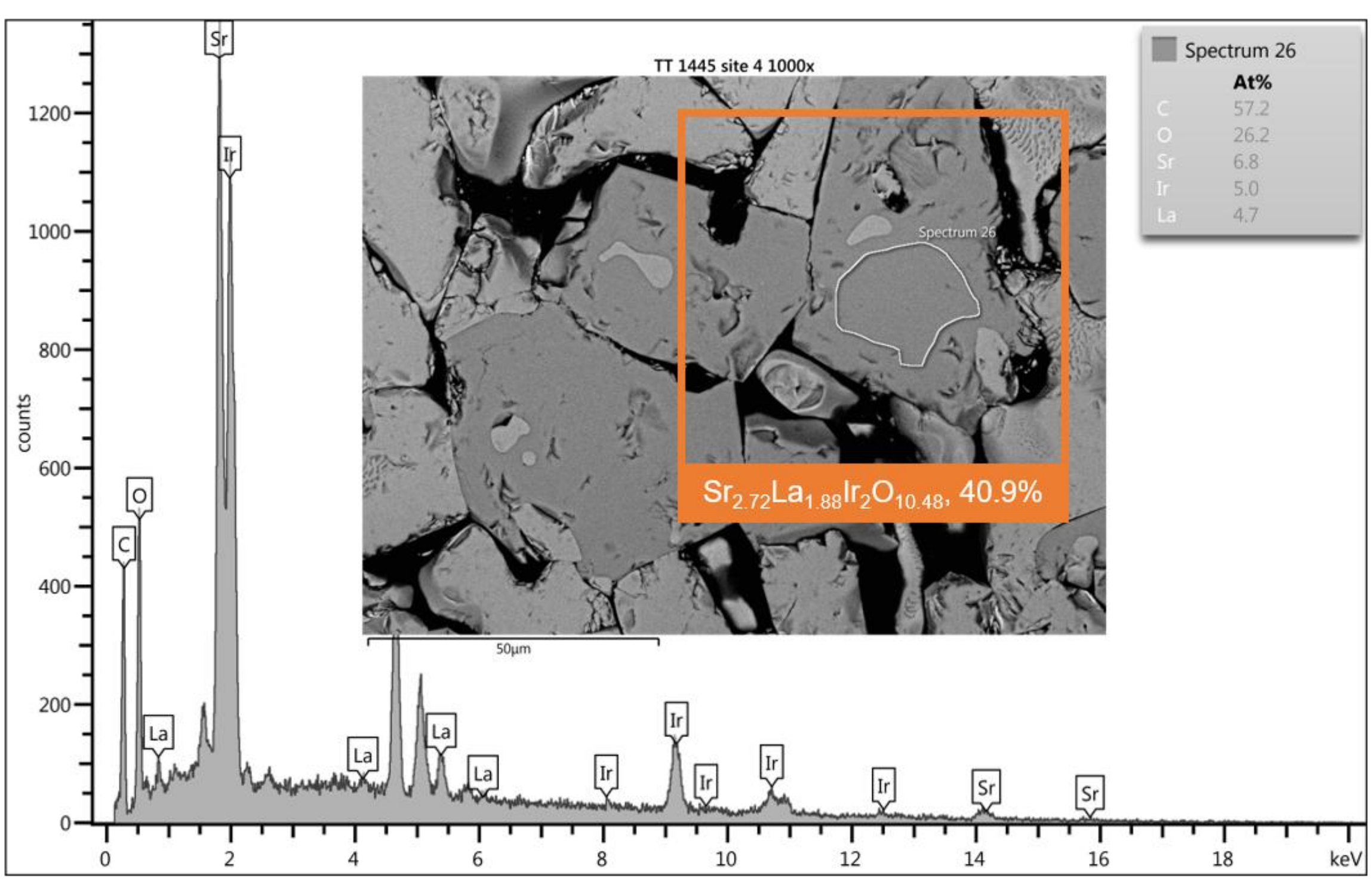

TT 1445 site 4 1000x
Spectrum 26
At%
C 57.2
O 26.2
Sr 6.8
Ir 5.0
La 4.7
Spectrum 26
$Sr_{2.72}La_{1.88}Ir_2O_{10.48}$, 40.9%
50µm
counts
keV